\documentstyle[aps,pre,twocolumn,epsf,floats]{revtex}
\begin{document}
\def\OP {\tensor P}
\def\B.#1{{\bbox{#1}}}
\renewcommand{\thesection}{\arabic{section}}
\title{{\sl DRAFT}  
{\rm \hfill  Version of \today}\\~~\\
  Computing the Scaling Exponents in Fluid Turbulence from First
  Principles:\\ Demonstration of Multi-scaling}
\author {Victor I. Belinicher$~^*$, Victor S. L'vov and Itamar  Procaccia}
\address{Department of~~Chemical Physics, The Weizmann
  Institute of Science, Rehovot 76100, Israel\\
$~^*$ Institute for Semiconductor Physics, Novosibirsk Russia.}
\maketitle
\begin{abstract}
This manuscript is a draft of work in progress, meant for network
distribution only. It will be updated to a formal preprint when the
numerical calculations will be accomplished. In this draft we develop 
a consistent closure procedure for the calculation of the scaling 
exponents $\zeta_n$ of the $n$th order correlation functions in fully
developed hydrodynamic turbulence, starting from first principles.
The closure procedure is constructed to respect the fundamental
rescaling symmetry of the Euler equation. The starting point of the 
procedure is an infinite hierarchy of coupled equations that are obeyed
identically with respect to scaling for any set of scaling exponents 
$\zeta_n$. This hierarchy was discussed in detail in a recent
publication  [V.S. L'vov and I. Procaccia, Phys. Rev. E, submitted, 
chao-dyn/970507015]. The scaling exponents in this set of equations
{\em cannot} be found from power counting. In this draft we discuss in
detail low order non-trivial closures of this infinite set of
equations,  and prove that these closures lead to the determination of
the scaling exponents from solvability conditions. The equations under
consideration after this closure are {\em nonlinear} integro-differential 
equations, reflecting the nonlinearity of the original Navier-Stokes 
equations. Nevertheless they have a very special structure such that
the  determination of the scaling exponents requires a procedure that
is  very similar to the solution of {\em linear} homogeneous equations, 
in which amplitudes are determined by fitting to the boundary conditions
in the space of scales. The re-normalization scale that is necessary for
any anomalous scaling appears at this point. The H\"{o}lder inequalities 
on the scaling exponents select the renormalizaiton scale as the outer
scale of turbulence $L$. 
\end{abstract}
\section{Introduction}
Unquestionably, the calculation of the scaling exponents that characterize
correlation functions in turbulence is one of the coveted goals
of nonlinear statistical physics. In a recent paper \cite{97LP}, hereafter
referred
to as paper I, a formal scheme to achieve such a calculation was
outlined. In the present paper we discuss the first order steps of
this scheme, identify
the mechanism for anomalous scaling in turbulence, and discuss the
series of steps available if one wants to improve the numerical values
of the computed anomalous exponents. The main ideas how to achieve the
lowest order closure
without breaking the rescaling symmetry of the Euler equation are
formulated such that they
repeat essentially unchanged in any higher order closure; it seems that no
new ideas are necessary.

To set up the developments of this paper, we present now a short review of
some essential ideas and results. Firstly we stress that a dynamical
theory of the statistics of turbulence
calls for a convenient transformation of variables that removes the
effects of kinematic sweeping. In our work we use the Belinicher-L'vov
transformation \cite{87BL} in which the field
 ${\B.v}({\B.r}_0,t_0\vert {\B.r},t)$ is defined in terms of the Eulerian
velocity
 ${\B.u}({\B.r},t)$:
 \begin{equation}
 {\B.v}({\B.r}_0,t_0\vert {\B.r},t)\equiv {\B.u}\lbrack{\B.r}
 +\mbox{\boldmath$\rho$}
 ({\B.r}_0,t),t\rbrack \ , \label{a2}
 \end{equation}
 where
 \begin{equation}
 \mbox{\boldmath$\rho$}  ({\B.r}_0,t)
  =\int_{t_0}^{t} ds {\B.u }[{\B.r}_0 +\mbox{\boldmath$\rho$}({\B.
  r}_0,s) ,s] \ .
  \label{a3}
 \end{equation}
 Note that $\mbox{\boldmath$\rho$} ({\B.r}_0,t)$ is precisely the
 Lagrangian trajectory of a fluid particle that is positioned at ${\bf
   r}_0$ at time $t=t_0$. The field ${\B.v}({\bf
   r}_0,t_0\vert {\B.r},t)$ is simply the Eulerian field in the
frame of reference of a single chosen material point $\B.\rho(\B.r_0,t)$.
Next we define the field of velocity differences $\B.{\cal
   W}(\B.r_0,t_0|\B.r,\B.r',t)$:
 \begin{equation}
\B.{\cal W}(\B.r_0,t_0|\B.r,\B.r',t)\equiv
\B.v( {\B.r}_0,t_0| {\B.r},t)-
\B.v( {\B.r}_0,t_0| {\B.r}',t)  \ . \label{newBL}
\end{equation}
It was shown that the equation of motion of this field is independent
of $t_0$, and we can omit this label altogether.

The fundamental statistical quantities in our study are the
many-time, many-point, ``fully-unfused", $n$-rank-tensor
correlation function of the BL velocity differences $\B.{\cal
  W}_j\equiv \B.{\cal W} (\B.r _0|\B.r_j \B.r'_j, t_j)$. To simplify
the notation we choose the following short hand notation:
\begin{equation}
X_j\equiv \{\B.r_j,\B.r'_j,t_j\}, \  x_j
\equiv \{\B.r_j,t_j\}, \  \B.{\cal W}_j\equiv
\B.{\cal W}(X_j) \ , \label{short}
\end{equation}
\begin{equation}
\B.{\cal F}_n(\B.r_0|X_1 \dots X_n)
= \left< \B.{\cal W}_1 \dots \B.{\cal W}_n \right> \ .
\label{defFtime}
\end{equation}
By the term ``fully unfused" we mean that all the coordinates are
distinct and all the separations between them lie in the inertial
range.  In particular the 2nd order correlation function written
explicitly is
\begin{eqnarray}
 &&{\cal F}_2^{\alpha\beta}( {\B.r}_{0}| {\B.r}_1,
 {\B.r_1}',t_1;\B.r_2,\B.r'_2,t_2)
 \nonumber \\ &&
=\langle {\cal W}^{\alpha}( {\B.r}_0| {\B.r}_1,\B.r'_1,t_1)
{\cal W}^{\beta}( {\B.r}_{0} | {\B.r}_2,\B.r'_2,t_2)\rangle\ .
 \label{Fab}
\end{eqnarray}
In addition to the $n$-order correlation functions the statistical
theory calls for the introduction of a similar array of response or
Green's functions.  The most familiar is the 2nd order Green's
function $G^{\alpha\beta} ({\B.r}_{0} |X_1;x_2)$ defined by the
functional derivative
 \begin{equation}
 G ^{\alpha\beta}({\B.r}_{0} |X_1;x_2) =
 \left\langle { \delta {\cal W}^{\alpha}( {\B.r}_0|X_1)
 \over  \delta \phi^{\beta}(\B.r_0| x_2)}\right\rangle \ .
 \label{green}
 \end{equation}
 In stationary turbulence these quantities depend on $t_1-t_2$ only,
 and we can denote this time difference as $t$.

The simultaneous correlation function $\B.T_n$ is obtained from
$\B.{\cal F}_n$ when $t_1=t_2\dots=t_n$. In this limit one can use differences
of Eulerian velocities, $\B.w(\B.r,\B.r',t)\equiv \B.u(\B.r',t)-\B.u(\B.r,t)$
instead of BL- differences, the result is the same.
In statistically stationary turbulence the equal
time correlation function is time independent, and we denote it as
\begin{eqnarray}
&&{\B.T}_n({\B.r}_1,{\B.r}'_1;{\B.r}_2,{\B.r}'_2;
\dots;{\B.r}_n,{\B.r}'_n)
\nonumber \\
&=& \left< \B.w({\B.r}_1,{\B.r}'_1,t)
\B.w({\B.r}_2,{\B.r}'_2,t) \dots
\B.w({\B.r}_n,{\B.r}'_n,t) \right> \ .
\label{defF}
\end{eqnarray}
One expects that when all the separations $R_i\equiv|\B.r_i-\B.r'_i|$
are in the inertial range, $\eta\ll R_i\ll L$, the simultaneous
correlation function is scale invariant in the sense that
\begin{eqnarray}
&&{\B.T}_n(\lambda {\B.r}_1,\lambda{\B.r}'_1;\lambda{\B.r}_2,
\lambda{\B.r}'_2;
\dots;\lambda{\B.r}_n,\lambda{\B.r}'_n)
\nonumber \\
&=& \lambda^{\zeta_n}{\B.T}_n({\B.r}_1,{\B.r}'_1;{\B.r}_2,
{\B.r}'_2;
\dots;{\B.r}_n,{\B.r}'_n) \ . \label{scaleinv}
\end{eqnarray}
The goal of our work is: {\em the calculation of the exponents $\zeta_n$
from first principles. This is first aim of a statistical
theory of turbulence}.

One could assume that also the time correlation
functions $\B.{\cal F}_n$ are homogeneous functions of their arguments,
including the time arguments. It should be stressed that this is not the
case, and that in the context of turbulence, when the exponents $\zeta_n$
are anomalous, dynamical scaling is broken. The time correlation functions
are ``multi-scaling" in the time variables. In Ref.~\cite{97LPP} it was
shown
that the scaling properties of the time correlation functions can be
parametrized conveniently with the help of one scalar 
function ${\cal Z}(h)$.
To understand this presentation, consider first the
simultaneous function $ T_n(\B.r_1\dots \B.r'_n)$.  Following
the standard ideas of multi-fractals \cite{95Fri,86HJKPS} the
simultaneous function can be represented as
\begin{eqnarray}
&&\B.T_n(\B.r_1,\B.r'_1\dots \B.r'_n)= U^n\int_{h_{\rm min}}
^{h_{\rm max}}
d\mu(h)\left({R_n\over L}\right)^{nh+{\cal Z}(h)}\nonumber \\ &&
\times \tilde\B.T_{n,h}(\B.\rho_1,\B.\rho'_1,\dots,\B.\rho'_n)
\ , \label{repres0}
\end{eqnarray}
 where $U$ is a typical velocity scale, and Greek coordinates stand for
dimensionless (rescaled) coordinates, i.e.
\begin{equation}
\B.\rho_j=\B.r_j/R_n \ , \quad \B.\rho'_j=\B.r'_j/R_n \ . \
\end{equation}
In Eq.~(\ref{repres0}) we have introduced the ``typical scale
of separation" of the set of coordinates
\begin{equation}
R^2_n={1\over n}\sum_{j=1}^n |\B.r_j-\B.r_j^{\prime}|^2 \ . \label{radgyr}
\end{equation}
At this point $L$ is an undetermined renormalization scale. At the end
of the calculation we will find that it is the outer scale of turbulence.
The function ${\cal Z}(h)$ is defined as
\begin{equation}
{\cal Z}(h) \equiv 3- {\cal D}(h)\ . \
\end{equation}
The function ${\cal Z}(h)$ is related to the scaling exponents
$\zeta_n$ via the saddle point requirement
\begin{equation}
\zeta_n = \min_h[nh+ {\cal Z}(h)] \ . \label{znzh}
\end{equation}
This identification stems from the fact that the integral in
(\ref{repres0}) is computed in the limit $R_n/L\to 0$ via the steepest
descent method. Neglecting logarithmic corrections one finds that
$\B.T_n\propto R_n^{\zeta_n}$.

The physical intuition behind the representation (\ref{repres0}) is
that there are velocity field configurations that are characterized by
different scaling exponents $h$. For different orders $n$ the main
contribution comes from that value of $h$ that determines the position
of the saddle point in the integral (\ref{repres0}).
It is convenient to introduce a dimensional quantity $\B.T_{n,h}$
according to
\begin{eqnarray}
\B.T_{n,h}(\B.r_1,\B.r'_1,\dots,\B.r'_n)=U^n\left({R_n\over
L}\right)^{nh+{\cal Z}(h)}\nonumber\\
\tilde T_{n,h}(\B.\rho_1,\B.\rho'_1,\dots,\B.\rho'_n) \ . \label{deftildeT}
\end{eqnarray}

Dimensional quantities of this type will play an important role
in our theory. Especially their rescaling properties will be used to
organize a ${\cal Z}$-covariant theory, see Section 3. This quantity
rescales like
\begin{equation}
\B.T_{n,h}(\lambda\B.r_1,\lambda\B.r'_1,\dots,\lambda\B.r'_n)=\lambda^{nh+
{\cal Z}(h)}
\B.T_{n,h}(\B.r_1,\B.r'_1,\dots,\B.r'_n) \ .
\end{equation}
Below we will use a shorthand notation for such rescaling laws, $\B.T_{n,h}\to
\lambda^{nh+{\cal Z}(h)} \B.T_{n,h}$.

The intuition behind the representation (\ref{repres0}) is extended to the time
domain. The particular velocity configurations that are characterized
by an exponent $h$ also display a typical time scale $t_{R,h}$ which
is written as
\begin{equation}
t_{R,h} \sim {R\over U} \left({L\over R}\right)^h \ .
\end{equation}
Accordingly we chose a temporal multi-scaling representation for
the time dependent function
\begin{eqnarray}
&&\B.{\cal F}_n(\B.r_0|X_1,\dots ,X_n)= U^n
\int\limits_{h_{\rm min}}^{h_{\rm max}}
d\mu(h)\left({R_n\over L}\right)^{nh+{\cal Z}(h)}\nonumber\\ &&
\times \tilde \B.{\cal F}_{n,h}(\B.r_0|\Xi_1,\Xi_2,\dots \Xi_n) \ ,
\label{repres}
\end{eqnarray}
where $\tilde\B.{\cal F}_{n,h}$ depends on the dimensionless
(rescaled) space and time coordinates
\begin{equation}
\Xi_j \equiv (\B.\rho_j, \B.\rho'_j,\tau_j)\ ,
 \quad \tau_j=t_j/t_{R_n,h} \ .
\end{equation}
As before, we introduce the related dimensional quantity 
$\B.{\cal F}_{n,h}$ according to:
\begin{eqnarray}
\B.{\cal F}_n(\B.r_0|X_1,\dots ,X_n)&&=U^n \left({R_n\over L}
\right)^{nh+{\cal Z}(h)}
\nonumber \\ &&\times \tilde\B.{\cal F}_{n,h}(\B.r_0|\Xi_1,
\Xi_2,\dots \Xi_n)
\end{eqnarray}
Of course, we require that the function 
$\tilde\B.{\cal F}_{n,h}(\B.r_0|\Xi_1,
\dots \Xi_n)$ would be identical to $\tilde
T_{n,h}(\B.\rho_1,\B.\rho'_1,\dots,\B.\rho'_n)$ when its rescaled time
arguments are all the same.
This representation reproduces all the scaling
laws that were found for time integrations and 
differentiations \cite{97LPP}. We
stress that when the multi-scaling representation of turbulent
structure functions was introduced for
the first time (\cite{95Fri} and references therein)
it was a phenomenological idea, that could
be understood as the inversion of the data
on $\zeta_n$, Eq.~(\ref{znzh}). We will show below that in the
context of our theory it appears as a result of an exact symmetry of the 
equations of motion. For our purposes it turns out easier
to compute the function ${\cal Z}(h)$ than to compute the exponents
$\zeta_n$ directly.

In paper I we showed that the $n$th order correlation functions satisfy, in the
limit of infinite Reynolds number, an
exact hierarchy of equation:
\FL
\begin{eqnarray}
&&{\partial\over \partial
t_1} \B.{\cal F}
_{n}(\B.r_0|X_1,\dots,X_n)\label{great}  \\&&+\int
d\tilde\B.r \B.\gamma(\B.r_1,\B.r_2,\tilde\B.r)
\B.{\cal F}_{n+1}(\B.r_0|\tilde X, \tilde X,X_2,
\dots X_n)=0 \ .\nonumber
\end{eqnarray}
The vertex function $\gamma^{\alpha\mu\sigma}(\B.r_1,\B.r'_1,
\tilde\B.r)$ is defined as
\begin{eqnarray}
&&\gamma^{\alpha\mu\sigma}(\B.r_1,\B.r'_1,
\tilde\B.r)={1\over 2}\{[P^{\alpha\mu}
(\B.r_1-\tilde \B.r)-P^{\alpha\mu}(\B.r'_1-
\tilde \B.r)]{\partial\over \partial \tilde
r_{\sigma}}\nonumber \\&&+[P^{\alpha\sigma}
(\B.r_1-\tilde \B.r)-P^{\alpha\sigma}(\B.r'_1
-\tilde \B.r)]{\partial\over \partial \tilde
r_{\mu}}\} \ . \label{gammanl}
\end{eqnarray}
The projection operator $P^{\alpha\beta}$ was defined
 explicitly in paper I.
It should be stressed at this point that the equations of motion do
not contain a dissipative term since we choose to deal with fully
unfused correlation functions. For such quantities the viscous 
term becomes negligible in the limit of Reynolds number Re$\to \infty$
or the viscosity $\nu\to 0$. This is the main advantage of working
with unfused correlation functions; the more commonly used
structure functions do not enjoy this property, and in their
equations of motion the viscous term remain relevant also in the
limit Re$\to \infty$. The price of working with unfused correlation
functions is that we have to deal with functions of many variables.

The aim of the present paper is to introduce a systematic method of solving
this chain of equations. We begin in section 2 with the introduction of the
central
idea that this chain of equations can solved on an ``$h$-slice". The set of
equations for the correlation functions has to be supplemented by
 an associated
chain of equations for the Green's functions, as is explained in Section 2.
In Section 3 we introduce the ${\cal Z}$-covariant closures which provide
a way to search solutions which are automatically scale invariant.
In Section 4 we explain in what sense ${\cal Z}(h)$ is obtained from
a solvability condition. In Section 5 we discuss what are the
missing calculational steps and summarize the present state of the
work.
\section{Fundamental equations on an ``$h$-slice"}
In this section we introduce equations on an ``$h$-slice" and consider
their symmetry under rescaling. Firstly we discuss the equations
for correlation functions, and secondly equations for Green's functions.
\subsection{Correlation functions and the rescaling group}
Examining the equations (\ref{great}) one realizes that they are
{\em linear} in the set of correlation functions $\B.{\cal F}_n$ with
$n=2,3,\dots$. In light of the representation (\ref{repres}) of
$\B.{\cal F}_n$ in terms of $\B.{\cal F}_{n,h}$  we can rewrite
Eqs. (\ref{great}) in the form
\FL
\begin{eqnarray}
&&\int_{h_{\rm min}}^{h_{\rm max}}
d\mu(h)\Big\{{\partial\over \partial
t_1} \B.{\cal F}
_{n,h}(\B.r_0|X_1,\dots,X_n)\label{great2}  \\&&+\int
d\tilde\B.r \B.\gamma(\B.r_1,\B.r_2,\tilde\B.r)
\B.{\cal F}_{n+1,h}(\B.r_0|\tilde X, \tilde X,X_2,
\dots X_n)\Big\}=0 \ .\nonumber
\end{eqnarray}
Since we integrate over a positive measure, the
equations are satisfied only if
the terms in curly parentheses vanish. In other words, we
will seek solutions to the equations
\begin{eqnarray}
&&{\partial\over \partial t_1} \B.{\cal F}_{n,h}(\B.r_0|X_1,\dots,X_n)+\int
d\tilde\B.r \B.\gamma(\B.r_1,\B.r_2,\tilde\B.r)\label{greater}
\\&&\times
\B.{\cal F}_{n+1,h}(\B.r_0|\tilde X, \tilde X,X_2, \dots X_n)= 0 \ . \nonumber
\end{eqnarray}
We refer to these equations as the equations on an ``h-slice". It is
important to analyze their properties under rescaling. To this aim recall that
the Euler equation is invariant to rescaling according to
\begin{equation}
\B.r\to \lambda \B.r \ , \quad t\to \lambda^{1-h}t \ ,
\quad \B.u \to \lambda^h \B.u \  ,  \label{ressym}
\end{equation}
for any value of $\lambda$ and $h$. On the basis of this alone
one could guess that that Eqs.~(\ref{greater}) are invariant under the
rescaling group
\begin{equation}
\B.r_i\to \lambda \B.r_i \ , \quad t_i\to \lambda^{1-h}t_i \ , \quad
\B.{\cal F}_{n,h}\to \lambda ^{nh} \B.{\cal F}_{n,h} \ . \label{wrongres}
\end{equation}
In fact we can see that Eqs.~(\ref{greater}) are invariant
to a broader rescaling group, i.e.
\begin{equation}
\B.r_i\to \lambda \B.r_i \ , \quad t_i\to \lambda^{1-h}t_i \ , \quad
\B.{\cal F}_{n,h}\to \lambda ^{nh+{\cal Z}(h)} \B.{\cal F}_{n,h} 
\ . \label{hslice}
\end{equation}
{\em with an $n$ independent function} ${\cal Z}(h)$. This extra freedom
is an exact result of the structure of the equations (\ref{greater}). It
is worthwhile to reiterate that this function appeared 
originally in the phenomenology
of turbulence as an ad-hoc model of multi-fractal turbulence. At this point
it appears as a nontrivial and {\em exact} property of the chain of of
equations of the statistical theory of turbulence.

We will show below that the preservation of this rescaling symmetry will
lead to a theory in which power counting plays no role.
As a consequence the information about the scaling
exponents $\zeta_n$ is obtainable only from the solvability conditions
of this equation. In other words, the information is buried in
coefficients rather than in power counting. The spatial derivative in
the vertex on the RHS brings down the unknown function ${\cal Z}(h)$,
and its calculation will be an integral part of the computation of the
exponents. It will serve the role of a generalized eigenvalue of the
theory.

Of course, we cannot consider the hierarchy of equations (\ref{greater})
in its entirety.  We need to find ways
to close this equation, and this is the main subject of section 3. The
main idea in choosing an appropriate closure is to preserve the
essential rescaling symmetry of the problem.  We will argue bellow
that there are many different possible closures, but most of
them do not preserve this rescaling symmetry. We will introduce the notion
of ${\cal Z}$ covariance, and demand that the closure does not destroy
the $h$-slice rescaling symmetry (\ref{hslice}).
\subsection{Temporal multi-scaling in the Green's functions}

In addition to the $n$-order correlation functions the statistical
theory calls for the introduction of a similar array of nonlinear response or
Green's functions $\B.{\cal G}_{m,n}$.
These represent the response of the direct product of $m$ BL-velocity
 differences to $n$ perturbations. In particular \FL
 \begin{eqnarray}
 &&\B.{\cal G}_{2,1}({\B.r}_0|X_1,X_2;x_3)=
 \left<{\delta [\B.{\cal W}({\B.r}_0|X_1)
  \B.{\cal W}({\B.r}_0|X_2)]\over \delta \B.\phi({\B.r}_0|x_3)}
\right>,\nonumber \\
 &&\B.{\cal G}_{3,1}({\B.r}_0|X_1,X_2,X_3;x_4)
 \nonumber \\ &&=\left<{\delta [\B.{\cal W}({\B.r}_0|X_1)
  \B.{\cal W}({\B.r}_0|X_2) \B.{\cal W}({\B.r}_0|X_3)]\over
 \delta\B.\phi({\B.r}_0|x_4)}
 \right>\ . \nonumber
 \end{eqnarray}
 Note that the Green's function $\B.G$ of Eq.~(\ref{green}) is
$\B.{\cal G}_{1,1}$ in this notation. The set of Green's functions
$\B.{\cal G}_{n,1}$ satisfies a hierarchy of equations that in the
limit of infinite Reynolds number is written as
\begin{eqnarray}
&&{\partial \over \partial t_1}{\cal G}
^{\alpha\beta\dots\psi\omega}_{n,1}
(\B.r_0|X_1,X_2,\dots,X_n;x_{n+1})
\nonumber\\&&+\int d\tilde \B.r\gamma^{\alpha\mu\sigma}
(\B.r_1,\B.r'_1, \tilde\B.r) \label{balG11}
\\&&\times{\cal G}^{\mu\sigma\beta\gamma
\dots\psi\omega}_{n+1,1}(\B.r_0\tilde X_1,\tilde X_1,
X_2,\dots,X_n;x_{n+1})\nonumber \\&&={\cal G}_{n,1}^{(0)
\alpha\beta\dots\omega}(\B.r_0|X_1,X_2\dots X_n,
\B.r_{n+1},t_1+0)\delta(t_1-t_{n+1})\ .\nonumber
\end{eqnarray}
The bare Green's function of $(n,1)$ order on the RHS of this equation
are the following decomposition:
\begin{eqnarray}
&&{\cal G}_{n,1}^{(0)\alpha\beta\dots\psi\omega}
(\B.r_0|X_1,X_2\dots X_n,
\B.r_{n+1},t_1+0)\nonumber \\&&\equiv G^{(0)
\alpha\omega}(\B.r_1,\B.r'_1,\B.r_{n+1})
{\cal F}^{\beta\gamma\dots\psi}_{n-1}(\B.r_0|X_2,
\dots , X_{n-1})\ . \label{Gn1}
\end{eqnarray}
Note that these functions serve as the initial conditions for 
Eqs.~(\ref{balG11})
at times $t_{n+1}=t_1$.

The form of these equations is very close to the hierarchy satisfied
by the correlation functions, and it is advantageous to use a similar
temporal multi-scaling representation for the nonlinear Green's functions:
\begin{eqnarray}
&&\B.{\cal G}_{n,1}
(\B.r_0|X_1,X_2,\dots,X_n;x_{n+1})=
\int\limits_{h_{\rm min}}^{h_{\rm max}}
d\mu(h)\nonumber\\ &&
\times \B.{\cal G}_{n,1,h}(\B.r_0|X_1,X_2,\dots X_n;x_{n+1}) \ .
\label{represG}
\end{eqnarray}
>From the rescaling symmetry of
the Euler equation we could guess the rescaling properties
$\B.{\cal G}_{n,1,h}\to \lambda^{(n-1)h-3}\B.{\cal G}_{n,1,h}$
As before, the equations support a broader rescaling symmetry group,
\begin{equation}
\B.r\to \lambda\B.r, \quad t\to \lambda^{1-h}t, \quad
\B.{\cal G}_{n,1,h}\to \lambda^{(n-1)h-3+{\cal Z}_G(h)}\B.{\cal G}_{n,1,h}
\ . \label{Gslice}
\end{equation}
In fact the choice of the scalar function ${\cal Z}_G(h)$ is constrained by
the initial conditions on the Green's functions.
>From Eq.~(\ref{Gn1}) we learn that the Green's functions depend on ${\cal
Z}(h)$ via
the appearance of the correlation functions. This means that ${\cal Z}_G(h)$
and ${\cal Z}(h)$ are related. A simple calculation leads to the conclusions
that
\begin{equation}
{\cal Z}_G(h)={\cal Z}(h) \ .
\end{equation}

In this subsection we displayed explicitly only the hierarchy of equations
satisfied
by $\B.{\cal G}_{n,1}$. Similar hierarchies can be derived for any family
of higher order Green's function $\B.{\cal G}_{m,n}$ with $m=2,3\dots$.
After introducing the multi-scaling representation, we can consider the
corresponding Green's functions on an ``$h$-slice", $\B.{\cal G}_{m,n,h}$
and show that the equations they satisfy have the rescaling symmetry
of the Euler equation with a ${\cal Z}(h)$ freedom. In all these equations
the initial value terms force the scalar function ${\cal Z}(h)$ to be
$m$-independent.
\section{${\cal Z}$-covariant closures}
Faced with infinite chains of equations, one needs to truncate at
a certain order. Simply by truncating one obtains a set of equations which
is not closed upon itself. It is then customary to express the highest
order quantities in the truncated set of equations in terms of lower
order quantities. This turns the set of equation into a nonlinear set.
Such an approach is not guaranteed to preserve the essential rescaling
symmetries (\ref{hslice}) and (\ref{Gslice}) of the equations.
In this section we develop a systematic method to close the hierarchies
of equations for correlation and Green's functions on an ``$h$-slice"
in a way that preserves the rescaling symmetry.

The lowest order closure involves five equations on an ``$h$-slice". The
first two belong to the $\B.{\cal F}_n$ hierarchy:
\begin{eqnarray}
&&{\partial\over \partial
t_1} \B.{\cal F}_{2,h}(X_1,X_2) +\int d\tilde\B.\rho \B.
\gamma(\B.r_1,\B.r_2,\tilde \B.r)
\label{2-3} \\ &&\times
\B.{\cal F}_{3,h}(\tilde X, \tilde X,X_2)=0 \ , \nonumber \\
&&{\partial\over \partial
t_1}\B.{\cal F}
_{3,h}(X_1,X_2,X_3) +\int d\tilde\B.r \B.
\gamma(\B.r_1,\B.r_2,\tilde\B.r)
\label{3-4} \\ &&\times
\B.{\cal F}_{4,h}(\tilde X, \tilde X,X_2,X_3)=0 \ . \nonumber
\end{eqnarray}
The next pair of equations belongs to the hierarchy of $\B.{\cal G}_{n,1}$.
Using the representation (\ref{represG}) in (\ref{balG11}) we derive:
\begin{eqnarray}
&&{\partial\over \partial
t_1} \B.{\cal G}
_{1,1,h}(X_1;x_2) +\int d\tilde\B.r \B.
\gamma(\B.r_1,\B.r_2,\tilde\B.r)
\label{g2-1} \\ &&\times
\B.{\cal G}_{2,1,h}(\tilde X, \tilde X;x_2)=\B.G^{(0)}_h(X_1;x_2)
\delta(t_1-t_2) \ . \nonumber \\
&&{\partial\over \partial t_1} \B.{\cal G}_{2,1,h}(X_1,X_2;x_3) +\int
d\tilde\B.r \B.
\gamma(\B.r_1,\B.r_2,\tilde\B.r)
\label{g2-3} \\ &&\times
\B.{\cal G}_{3,1,h}(\tilde X, \tilde X,X_2;x_3)=0 \ . \nonumber
\end{eqnarray}
Here $\B.G^{(0)}_h$ stands for the bare Green's function on an 
``$h$-slice".
The fifth needed equation is the first equation from the third hierarchy
of Green's functions $\B.{\cal G}_{n,2,h}$:
\begin{eqnarray}
&&{\partial\over \partial
t_1} \B.{\cal G}_{1,2,h}(X_1;x_2,x_3) +\int d\tilde\B.r \B.
\gamma(\B.r_1,\B.r_2,\tilde\B.r)
\label{g1-2} \\ &&\times
\B.{\cal G}_{2,2,h}(\tilde X, \tilde X;x_2,x_3)=0 \ . \nonumber
\end{eqnarray}


\begin{figure}
\epsfxsize=8.6truecm
\epsfbox{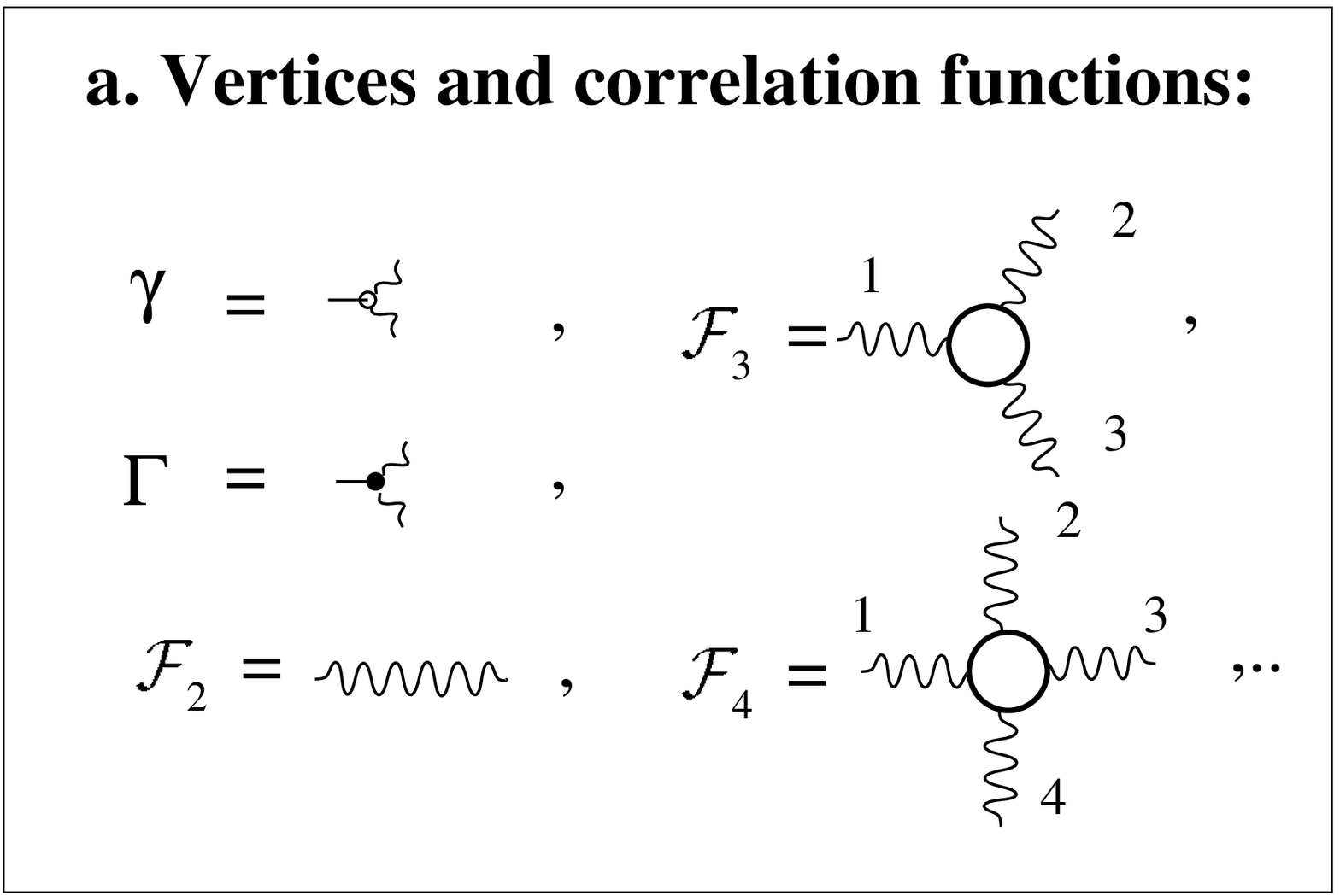}
\epsfxsize=8.6truecm
\epsfbox{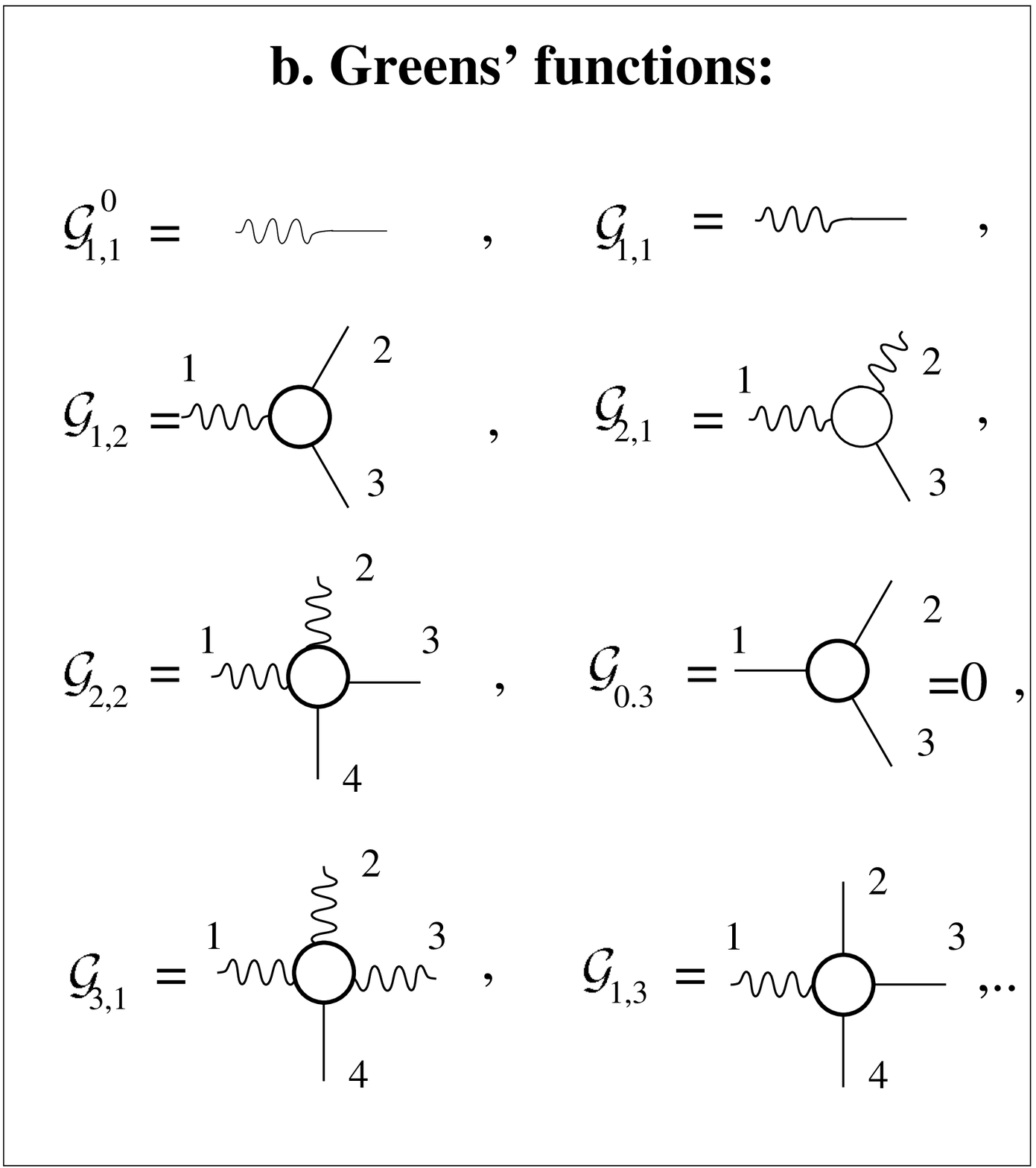}
\caption
{The diagrammatic notation of the basic objects of the theory.  Panel
  a: the vertex $\gamma$ and the correlation functions ${\cal F}_n$
  with $n=2,3,4$. 
Panel b: the bare Green's function
  ${\cal G}^0_{1,1}$ (thin line), and the dressed Green's functions
  ${\cal G}_{n,m}$. Objects with only straight tails are identically
  zero.}
\label{F0}
\end{figure}

\begin{figure}
\epsfxsize=8.6truecm
\epsfbox{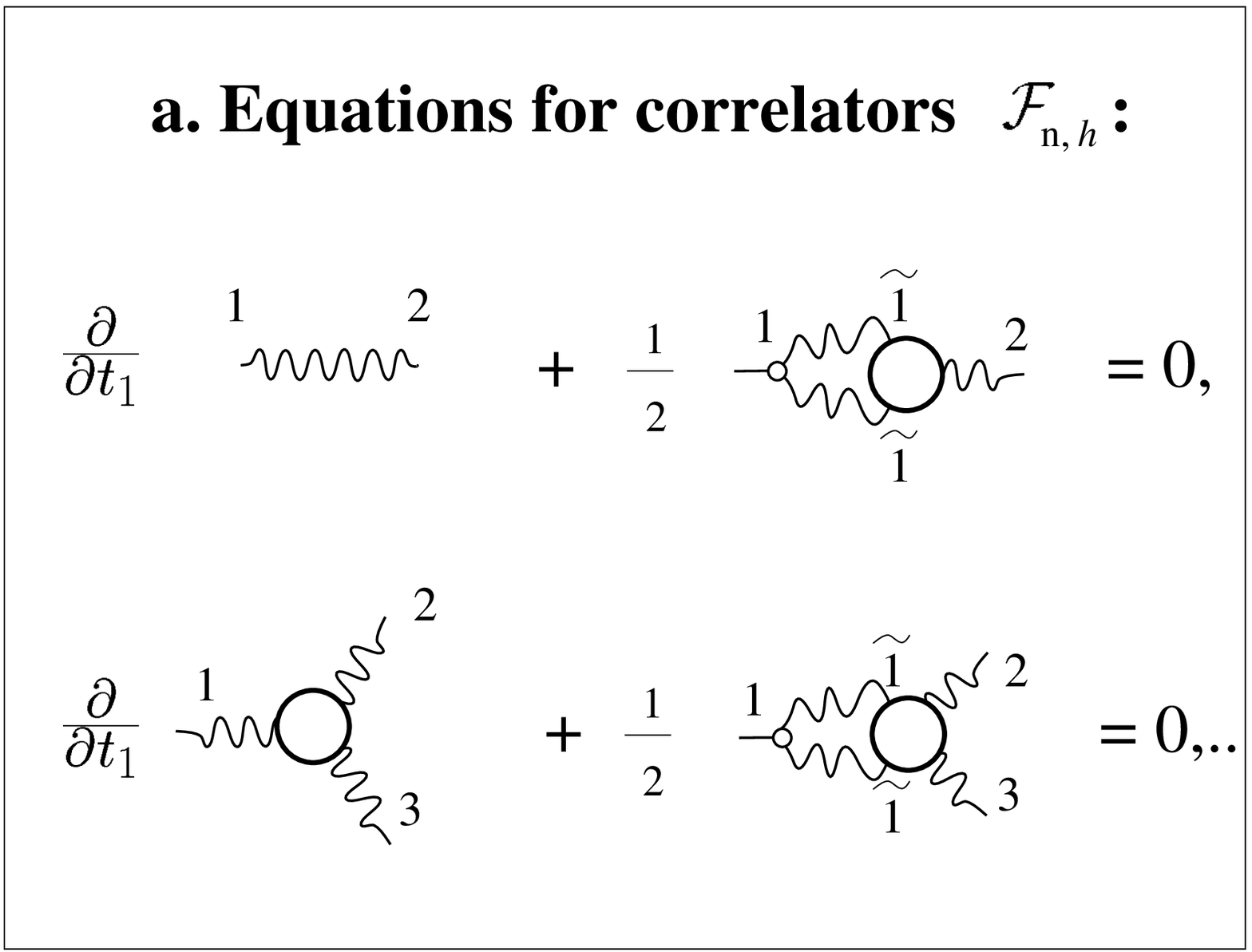}
\epsfxsize=8.6truecm
\epsfbox{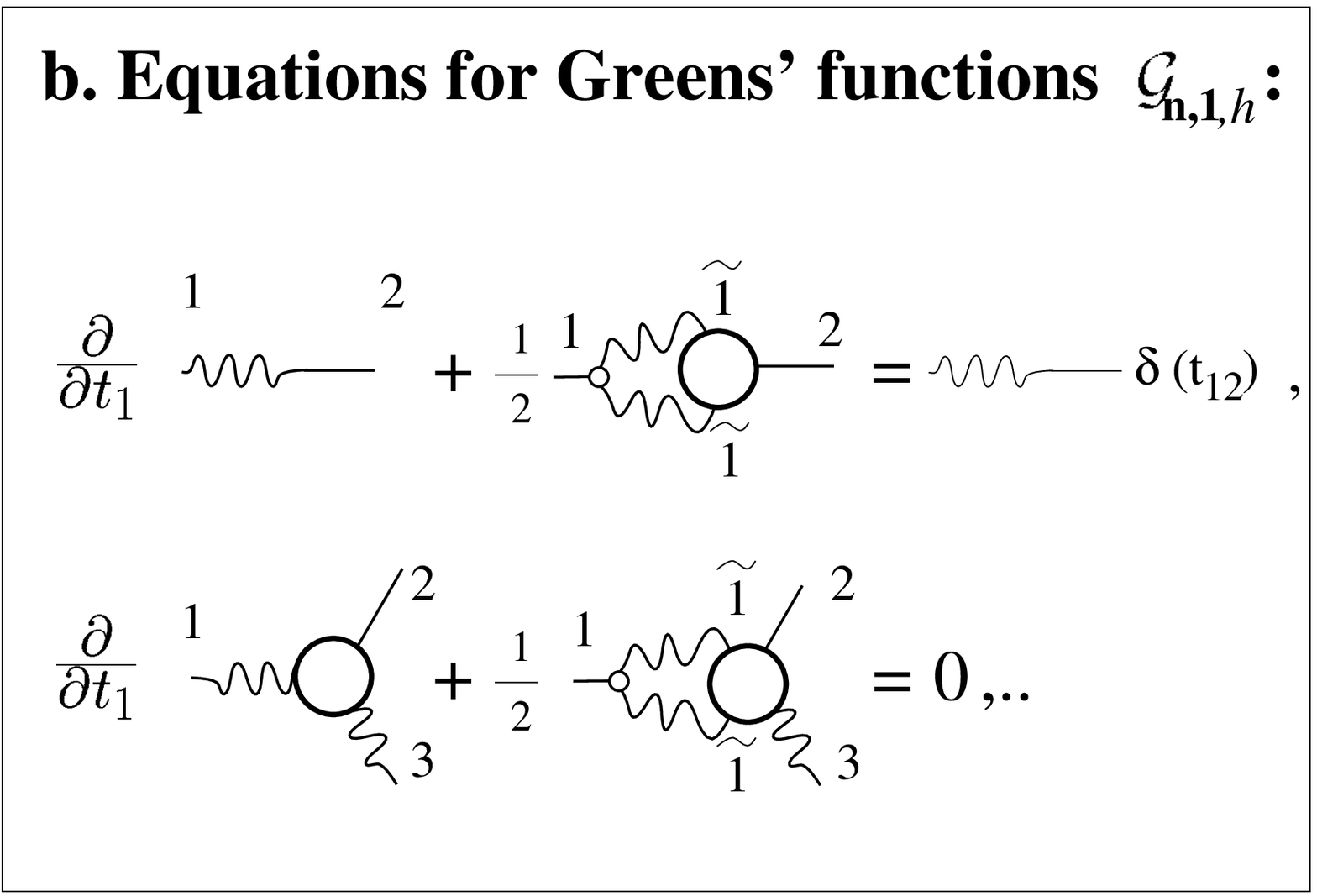}
\epsfxsize=8.85truecm
\epsfbox{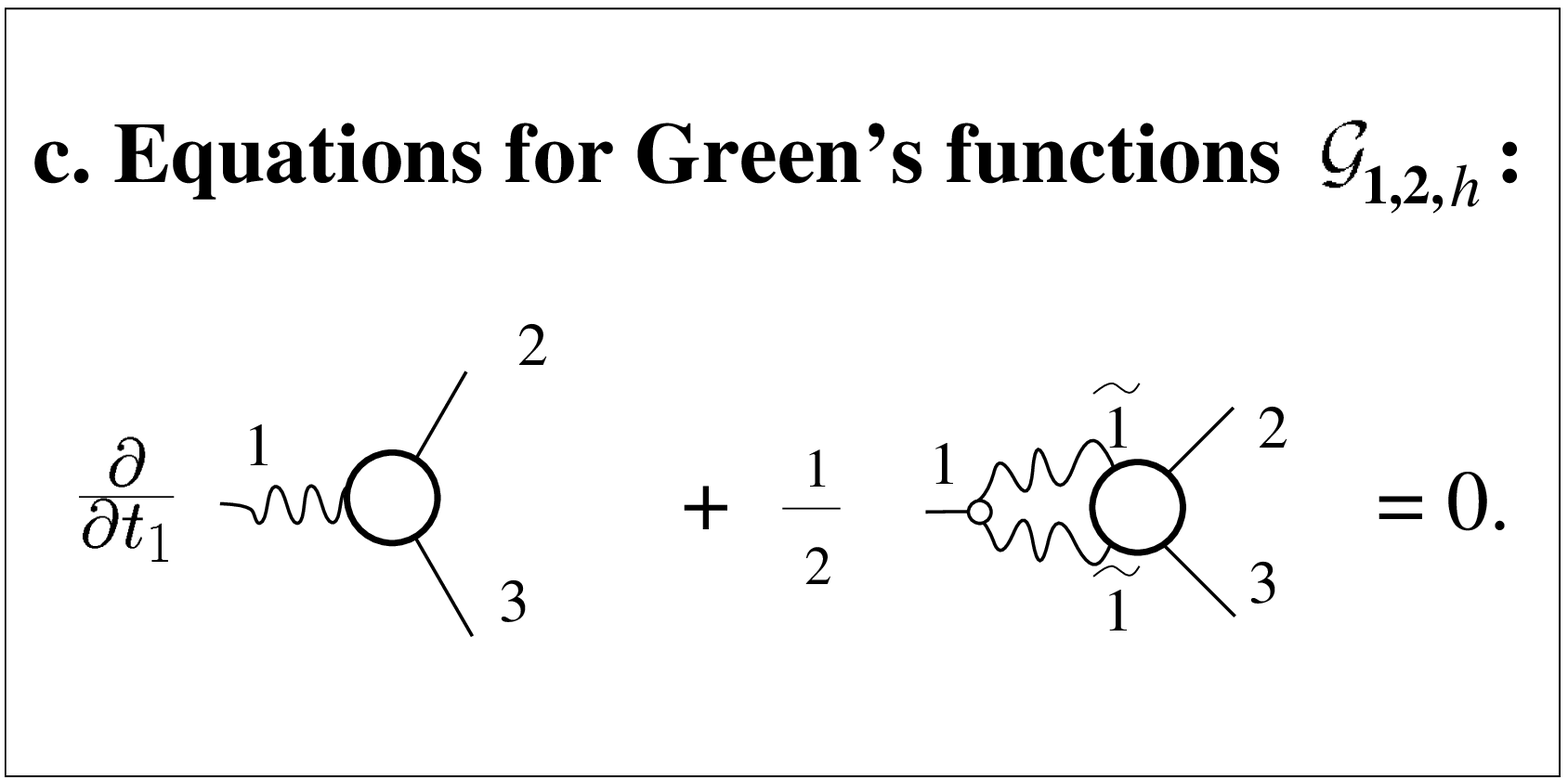}
\vskip 0.5cm 
\caption{Diagrammatic reprezentation of the five equations for the 
lowest nontrivial ${\cal Z}$-covarian closure. }
\label{F1}
\end{figure}

\begin{figure}
\epsfxsize=8.6truecm
\epsfbox{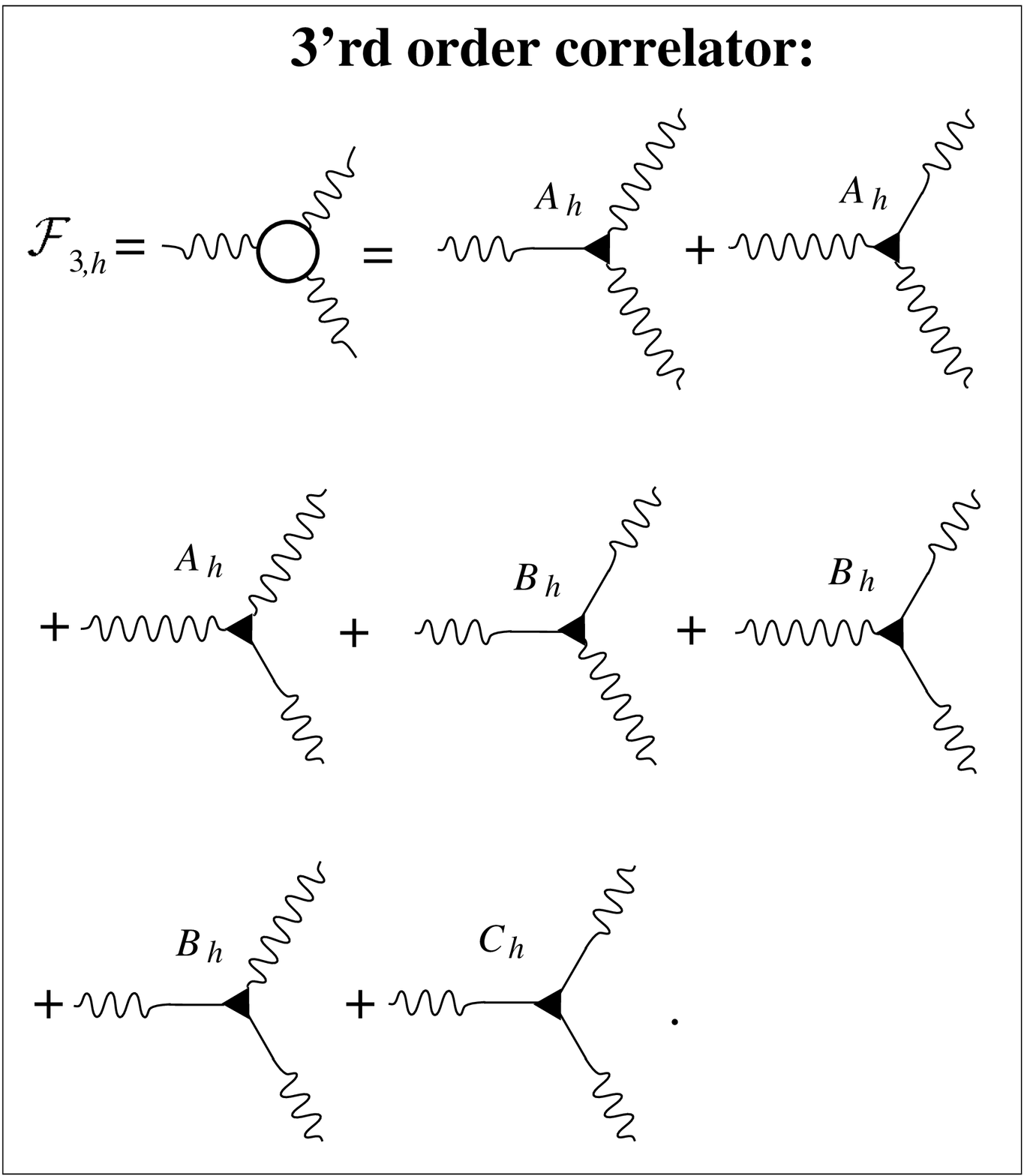}
\vskip 0.5cm 
\caption
{Exact reprezentation of the third-order correlation function    }
\label{F2}
\end{figure}
\begin{figure}
\epsfxsize=8.6truecm
\epsfbox{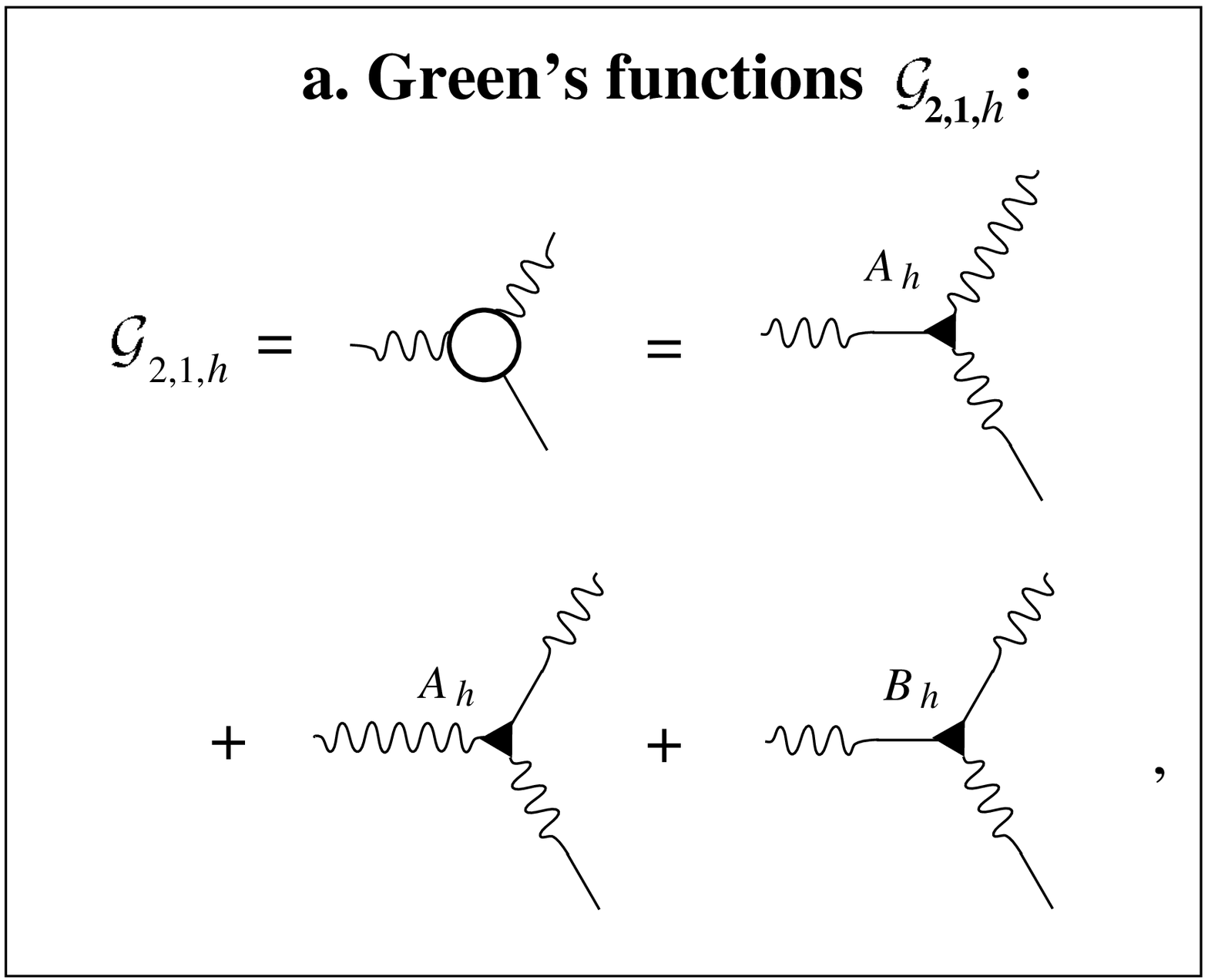}
\epsfxsize=8.6truecm
\epsfbox{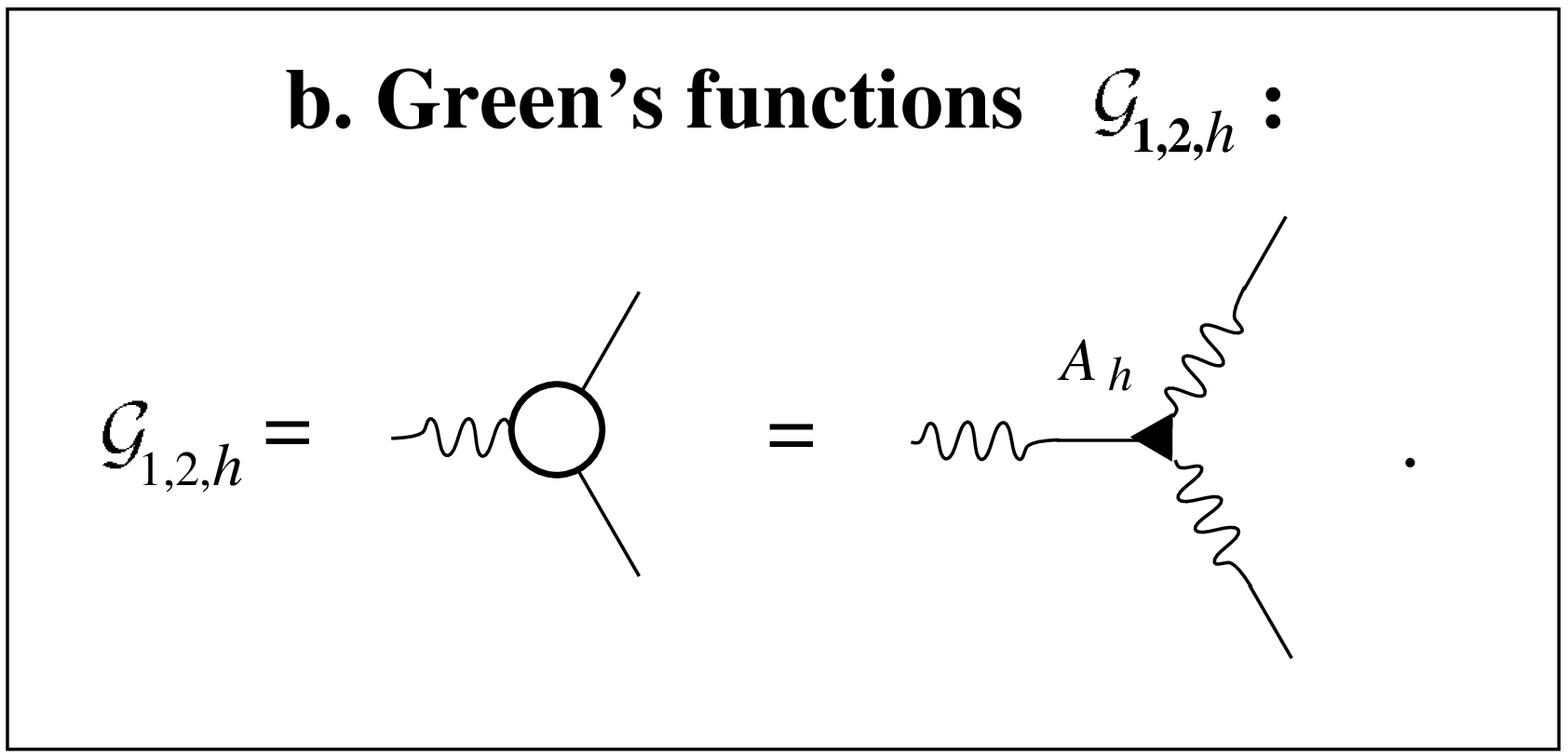}
\vskip 0.5cm   
\caption{  
Exact representation of the nonlinear Green's functions ${\cal
    G}_{2,1}$ and ${\cal G}_{1,2}.$  }   
\label{F3}
\end{figure}
These five equations are presented symbolically in Fig.~\ref{F1} 
and the symbols are
explained in Fig.~\ref{F0}. We now show that in the first step
of the closure procedure these five equations can be considered as
${\cal Z}(h)$-covariant equations for five unknowns. These five objects
are the 2nd order correlation function $\B.{\cal F}_{2,h}$, the regular
Green's function $\B.{\cal G}_{1,1,h}$, and three types of triple vertices.
The vertices are introduced in Figs.~\ref{F2} and \ref{F3}. 
We have in these
figures three relationships that {\em define} the vertices $\B.A_h$,
$\B.B_h$ and $\B.C_h$ {\em on an ``h-slice"} in terms of 
$\B.{\cal F}_{3,h}$, $\B.{\cal G}_{2,1,h}$
and $\B.{\cal G}_{1,2,h}$. Note that there is no notion of 
perturbation theory
here - we simply define the three vertices in terms of objects that appear
in Eqs. (\ref{2-3}),(\ref{g1-2}) and (\ref{g2-1}).

Eq.~(\ref{3-4}) involves the 4th point correlator $\B.{\cal F}_{4,h}$,
Eq.~(\ref{g2-3})
and Eq.~(\ref{g1-2}) involve 
$\B.{\cal G}_{3,1,h}$ and $\B.{\cal G}_{2,2,h}$.
These are 4th order objects, and we present
them in Figs.~\ref{F4} terms of all the possible decompositions made of lower
order objects, and in addition new (``irreducible") contributions which
{\em are defined} by these relations. In order to have a consistent
definition we need to add to this game the Green's function $\B.{\cal
G}_{1,3,h}$.
In the context of the 4th order objects
the irreducible contributions are denoted symbolically
as empty squares. There are four of them, and we denote them as
$\B.D_{3,1,h}$, $\B.D_{2,2,h}$, $\B.D_{1,3,h}$ and $\B.D_{0,4,h}$.
 The first
index stands for the number of wavy ``tails" and the second index
 for the number
of straight tails of the empty square.
\subsection{Systematic Closure}
{\em The first step of closure consists of discarding the irreducible empty
squares}. After doing so, we remain with precisely five equations (\ref{2-3})--
(\ref{g1-2}) for five unknown functions. In the next step of closure we
retain the empty squares as defined by their relations to the 4th order
correlation and Green's function, and {\em add to the list of equations
on an ``h-slice" the equations of motion for the 4th order objects}, i.e.
$\B.{\cal F}_{4,h}$, $\B.{\cal G}_{3,1,h}$, $\B.{\cal G}_{2,2,h}$ and
$\B.{\cal G}_{1,3,h}$. In total we have at this point
nine equations. These equations will involve four 5th order objects,
i.e. $\B.{\cal F}_{5,h}$, $\B.{\cal G}_{4,1,h}$, $\B.{\cal G}_{3,2,h}$
and $\B.{\cal G}_{2,3,h}$. Each of these new objects can be written now
in terms of all the contributions that can be made from low order objects,
plus irreducible 5th order vertices that we denote as empty pentagons.
The second step of closure consists of discarding the empty pentagons.
This gives us precisely nine equations for nine unknown functions, i.e.
$\B.{\cal F}_{2,h}$, $\B.{\cal G}_{1,1,h}$, $\B.A_h$, $\B.B_h$, $\B.C_h$,
and the four empty square vertices $\B.D_{3,1,h}$, $\B.D_{2,2,h}$,
$\B.D_{1,3,h}$
and $\B.D_{0,4,h}$.

The procedure is now clear in its entirety. At the $n$th step of the
 closure
we will discard the $n+3$th irreducible contributions, and will have
precisely the right number of equations on an ``$h$-slice" to solve for the
remaining unknowns. We should stress that this procedure is not 
perturbative
since we solve the exact equations on an ``$h$-slice". Our presentation
of $n$th order objects on the ``$h$-slice" in terms of lower order ones is
also exact, it just {\em defines} at the $n$th step of the procedure
a group of $(n+3)$th new vertices. These vertices are solved for only at
the $(n+1)$th step of the procedure, when the $(n+4)$th vertices are
 discarded.
\begin{figure}
\epsfxsize=8.6truecm
\epsfbox{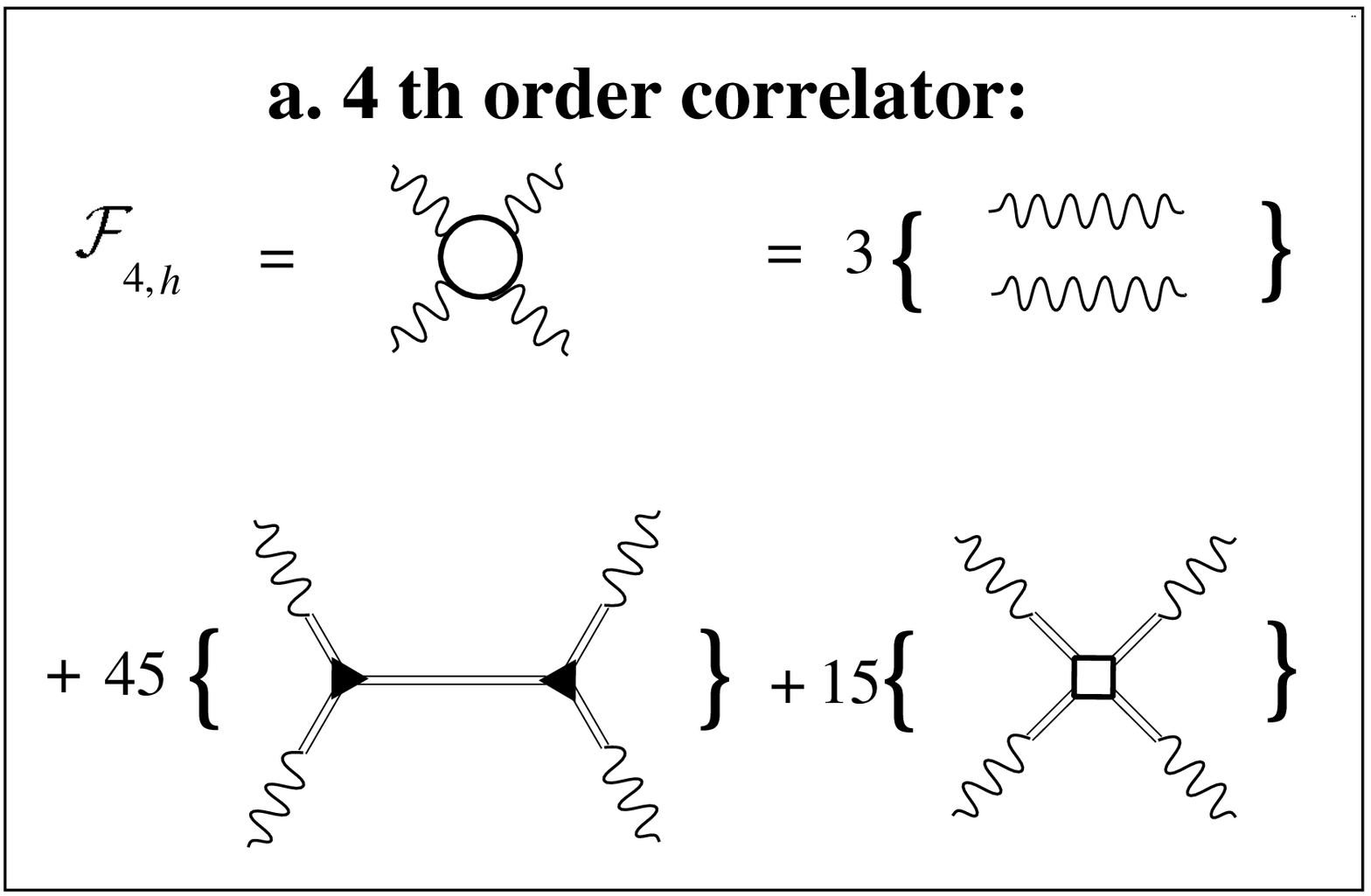}\vskip 0.1cm
\epsfxsize=8.6truecm
\epsfbox{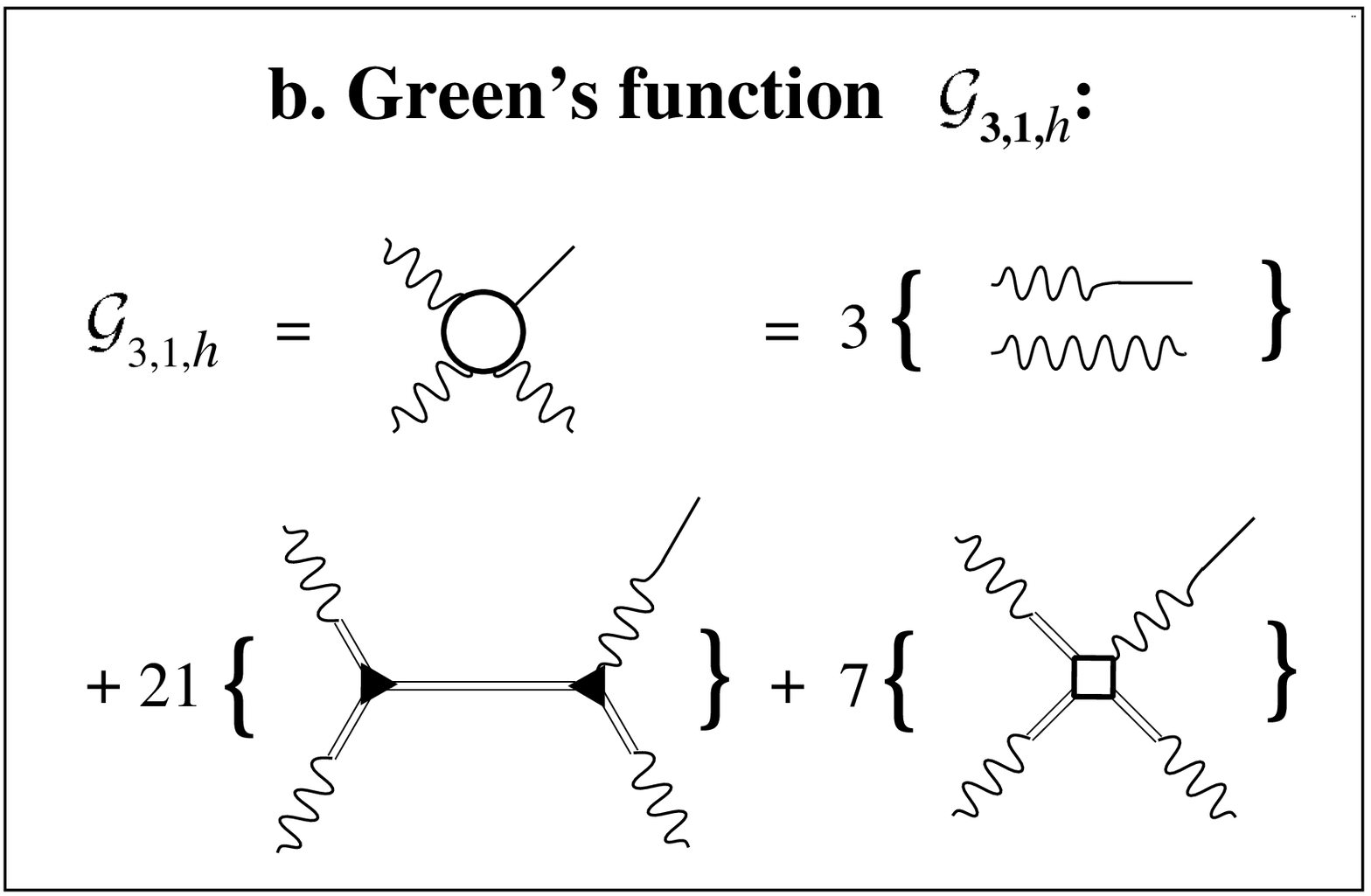}\vskip 0.1cm
\epsfxsize=8.6truecm
\epsfbox{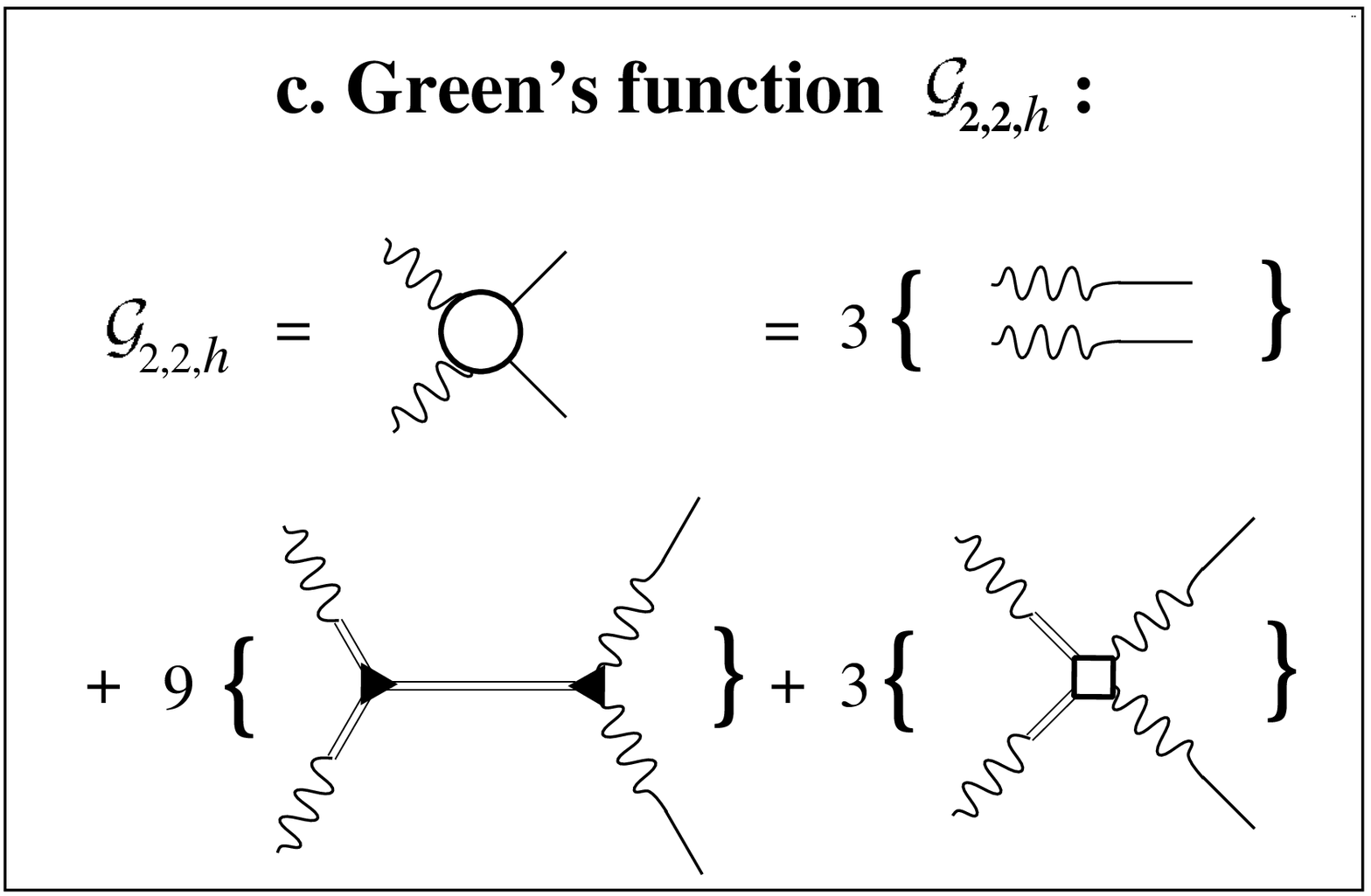}
\vskip 0.5cm 
\caption{
Exact representation of the forth-oredr correlation function
(Panel a) and  nonlinear Green's functions (Panels b and c).
Empty  squares reprezent irreducible contributions to the 4th
order vertices which are neglected in the lowest order ${\cal
  Z}$-covariant closure. Double line represent either wavy or
straight line. For more details see Paper I.}  
\label{F4}
\end{figure}
It would be only fair to say that the idea for this procedure came from
a careful examination of the fully renormalized perturbation 
theory for this problem, \cite{97LP}. In that procedure one can {\em derive} equations
for the $nth$ order objects that appear symbolically like
Figs.~\ref{F4}a, \ref{F4}b. etc.
In addition, one can at each step of the procedure have an infinite
expansion for the irreducible contributions. Nevertheless, the procedure
explained above is different; firstly, it is consistently developed on
an ``$h$-slice", whereas the renormalized perturbation theory is done
for the standard statistical objects. Secondly, at no point is there
any infinite expansion whose convergence properties are dubious. We just
go through a set of explicit definitions solving an exact set of equations.
The only question that needs to be understood is the speed of convergence
of this scheme in terms of the scalar function ${\cal Z}(h)$ which
parametrizes the anomalous behavior.
\subsection{${\cal Z}$-covariance}
A crucial property of our closure procedure is that it guarantees
that power counting remains irrelevant on an ``$h$-slice" for an
arbitrary step of the procedure, and the scalar function ${\cal Z}(h)$
cannot be computed from power counting. To see this we need to find
the rescaling properties of the triple and higher order vertices.
We start with the triple vertices $\B.A_h$, $\B.B_h$ and $\B.C_h$.
The first one is defined by its relation to $\B.{\cal G}_{1,2,h}$,
see Fig.~\ref{F3}b. Using the facts that
\begin{eqnarray}
\B.{\cal F}_{2,h}&\to&\lambda^{2h+{\cal Z}(h)} \B.{\cal F}_{2,h}\ ,
\label{rescaleF2}\\
\B.{\cal G}_{1,1,h}&\to&\lambda^{{\cal Z}(h)-3}\B.{\cal G}_{1,1,h}\ ,
\label{rescaleG11}\\
\B.{\cal G}_{1,2,h}&\to&\lambda^{-h+{\cal Z}(h)-6}\B.{\cal G}_{1,2,h}\ ,
\label{rescaleG12}.
\end{eqnarray}
we find that $\B.A_h$ has to transform according to
\begin{equation}
\B.A_h \to \lambda^{2h-2{\cal Z}(h)-9}\B.A_h \ . \label{rescaleA}
\end{equation}
Armed with this knowledge we proceed to the definition of $\B.B_h$ through
its relation to $\B.{\cal G}_{2,1,h}$, equation in Fig.~\ref{F3}a. 
We can check
that the rescaling property of
\begin{equation}
\B.{\cal G}_{2,1,h}\to\lambda^{h+{\cal Z}(h)-3}\B.{\cal G}_{1,2,h}
\end{equation}
agrees exactly with the two terms that contain the vertex $\B.A_h$ on the
RHS. Accordingly also the term containing $\B.B_h$ has to transform
in the same way, leading to
\begin{equation}
\B.B_h \to \lambda^{4h-2{\cal Z}(h)-6}\B.B_h \ . \label{rescaleB}
\end{equation}
In making this assertion we assumed that there is no cancellation of the
leading terms in the equation. Otherwise the vertex $\B.B_h$ would be
smaller.
Lastly we use the definition of $\B.C_h$ by the relation to $\B.{\cal F}_{3,h}$,
see Fig.~\ref{F2},
that transforms like
\begin{equation}
\B.{\cal F}_{3,h}\to\lambda^{3h+{\cal Z}(h)} \B.{\cal
F}_{3,h}\label{rescaleF3} \ .
\end{equation}
All the terms that include $\B.A_h$ and $\B.B_h$ have the same rescaling
exponent as that of $ \B.{\cal F}_{3,h}$. We can therefore find the rescaling
exponent of $\B.C_h$:
\begin{equation}
\B.C_h \to \lambda^{6h-2{\cal Z}(h)-3}\B.C_h \ . \label{rescaleC}
\end{equation}
We again assumed that there is no cancellation of the leading
terms in the equation for $\B.{\cal F}_{3,h}$. If there is cancellation,
the vertex $\B.C_h$ can be smaller.

At this point we can check that the first step in our closure scheme
leads to a ${\cal Z}(h)$ covariant procedure. Consider firstly
the three contributions of Gaussian decomposition which are first
on the RHS of the equation in Fig.~\ref{F4}a. These rescale like
$\lambda^{4h+2{\cal Z}(h)}$,
and their ratio to the LHS is proportional to $(R/L)^{{\cal Z}(h)}$. For
${\cal Z}(h)$ positive these contributions become irrelevant in the limit
$(R/L)\to 0$. The 45 contributions that come next contain pairs of
triple vertices, and we need to use
the rescaling properties (\ref{rescaleA}), (\ref{rescaleB}), and
(\ref{rescaleC})
to find their rescaling exponents. We find that they 
{\em all share the same rescaling exponent} . In hindsight this
should not be surprising. This is a result of the {\em assumption} that
in the definitions of the three vertices there are no cancellations
in the leading scaling behavior. Thus the rescaling exponent of all the
45 contributions could be obtained from analyzing
 one of them. The nontrivial
fact is that the common rescaling of all these terms is exactly
the rescaling of the LHS of the equation, which is
$\lambda^{4h+{\cal Z}(h)}$ . This means that our closure for the 4th order
correlation functions cannot introduce
power counting. Note that the rescaling neutrality with respect to counting
of $h$ and of natural numbers follows from the rescaling symmetry of the
Euler equation, and is shared also by Gaussian contributions. On the other hand
the neutrality with respect to
 ${\cal Z}(h)$ is nontrivial, and follows from
a judicious choice of the proposed closure scheme. We will refer this
property 
as ${\cal Z}$-covariance.

It is important to understand now that the proposed closures
for the other 4th order objects, like Fig.~\ref{F4}b
 are also ${\cal Z}$-covariant.
All that changes
is the number of wavy and straight tails on the LHS and RHS of the equations,
and the rescaling exponents change in the same way on the two sides of
the equation.

We can now consider the next step of closure, taking into account the
irreducible 4th order vertices (empty squares), discarding the 5th
order empty pentagons (irreducible contributions, for details, see
Paper 1  The procedure follows
verbatim the one described above for the triple vertices and 
the rescaling
exponents of the irreducible 4th order vertices $\B.D_{m,n,h}$ are determined:
\begin{eqnarray}
\B.D_{m,n,h}&\to& \lambda^{d_{m,n}(h)}\B.D_{m,n,h} \ , \label{rescaleD}\\
d_{m,n}(h)&=&2nh-3{\cal Z}(h)-3m-4 \ . \nonumber
\end{eqnarray}
We can check the terms that appear in the 5th order correlation
and Green's functions,
which are made of combinations of triple and 4th order vertices.
We discover that all these terms share the same rescaling
exponent, and that it agrees precisely with the rescaling of the
5th order correlation and Green's function. Accordingly also the second
step of closure is ${\cal Z}$-covariant.

It becomes evident that we develop a systematic ${\cal Z}$-covariant closure
scheme, and that power counting will not creep in at any step of the
procedure. In the next section we show that the scalar function ${\cal Z}(h)$
can be computed from this scheme as a generalized eigenvalue. We also explain
the role of the boundary conditions in the space of scales, and how the
re-normalization scale is chosen.
\section{The scalar function ${\cal Z}$ as a generalized eigenvalue}

In this section we demonstrate that at any step of our closure, the scalar
function
${\cal Z}(h)$ can be found only from a solvability condition. We will also
explain the role of the boundary conditions in the space of scales
in determining the scaling functions in this theory.

The point is really rather simple. First observe that our initial equations
for correlation and Green's functions on an ``$h$-slice", like (\ref{greater})
or (\ref{g2-1})-(\ref{g1-2}) are {\em linear} functional equations.
The equations for the correlation functions are not only linear, but
also homogeneous. The equations for the Green's functions are not all
homogeneous,
but as we explained in the last section the inhomogeneous terms are much
smaller than the homogeneous terms in the limit $(R/L)\to 0$, and they
can be discarded.  This means, of course, that all the functions on
an ``$h$-slice" can
be determined only up to an over all numerical constant. At this point
we may have even more than one overall free constant; on the face
of it the equations for the correlation functions are independent of
the equations for the Green's functions, and we can have different
overall constants in the correlation and the Green's functions.

Every step of
closure turns a set of linear functional equations into a set of
nonlinear functional equations. We claim that nevertheless all the functions
appearing in these equations can be determined only up to an overall
numerical constant. The extra freedom of many possible constants
disappears now, since the correlation and the Green's functions are
coupled after closure. But one overall constant remains free.
The reason is of course the property of ${\cal Z}$-covariance,
which includes as a special case invariance to an overall scaling factor,
or rescaling by $\lambda^0$.

If our equations were linear, this freedom would have meant that we
need to require the standard solvability condition that the linear operator
had an eigenvalue zero, and the functions that we seek would have been
identified as the ``zero-modes" associated with this eigenvalue. Our
equations are not linear, and the solvability condition is not that simple.
Nevertheless we know apriori that at every step of the closure we will need
to find the solvability condition of the nonlinear set of equations, and
this condition will determine the numerical value of ${\cal Z}(h)$.

Even after determining ${\cal Z}(h)$ the statistical functions will
be determined only up to a factor which may depend on $h$, which is
the measure $\mu(h)$ in Eqs(\ref{repres}) and (\ref{represG}).
In order to determine this factor we will need to fit all correlation
functions $\B.T_n$
to the boundary conditions in the space of scales. At that point the
computed values of ${\cal Z}(h)$ will determine whether it is the
inner or the outer scale of turbulence that appears as the
renormalizaiton scale. We will show in the lowest step of closure
that the outer scale $L$ is selected.
\section{Summary and the Road Ahead}
Up to now we described the general ideas and how the closure
scheme should work. The main point of the analysis is that the
equations of motion of the statistical objects admit an exact
rescaling group that contains, for a given ``$h$-slice", one 
unknown scalar function ${\cal Z}(h)$ whose calculation is
sufficient for the evaluation of the scaling exponents $\zeta_n$
via Eq.~(\ref{znzh}). 

Unfortunately, the actual calculations that are called for 
in this scheme are far from trivial. In the lowest step
of closure we are faced with five coupled integro-differential
equations in many space-time variables, and the complexity
increases rapidly in the higher steps of closure. We are 
currently attempting to solve numerically the lowest step
of closure with the aim of {\em demonstrating} the existence
of anomalous scaling. We are not particularly interested in
the first step in precise values of the exponents $\zeta_n$,
it is more important to show that they differ from their K41
counterparts. Numerical accuracy and the convergence of
${\cal Z}(h)$ will be examined in the higher steps of closure.
The complexity of the numerics means that this is a long
program that is expected to last for a couple of years.
For this reason we chose to describe the ideas of the
closure scheme even at this stage when the numerical implementation
is lacking. It is our feeling that the procedure is sound 
and that there is 
a good chance to obtain results that will justify the effort
that is called for in the numerical implementation.

\end{document}